\documentclass[options]{JHEP3}

\title{A Brane Teaser.}

\author{Tasneem Zehra Husain\\ 

Department of Physics\\ Stockholm University\\
PO Box 6730\\ S 11385 Stockholm\\ 
Sweden\\ Email: \email{tasneem@physto.se}}

\abstract{In this note we study the puzzle posed by two M5-branes intersecting on a string 
(or equivalently, a single M5-brane wrapping a holomorphic four-cycle in $\C^4$). It has been 
known for a while that this system is different from all other configurations built using
self-intersecting M-branes;  in particular the corresponding supergravity solution exhibits 
various curious features which have remained unexplained. We propose that the resolution 
to these puzzles  lies in the existence of a non-zero two-form on the M5-brane world-volume. }

\keywords{M5-branes, World-Volume Fields, Supergravity Solutions}

\preprint{USITP-03-06 \\ hep-th/0306248}

\newcommand{\C}{{\mathbb C}}

\newcommand{\comment}[1]{}

\def\bbbz{{\sf Z\!\!\!Z}}
\def\sl2z{SL(2,\bbbz)}

\newcommand{\be}{\begin{equation}}
\newcommand{\ee}{\end{equation}}
\newcommand{\bea}{\begin{eqnarray}}
\newcommand{\eea}{\end{eqnarray}}

\def\bbbz{{\sf Z\!\!\!Z}}
\def\sl2z{SL(2,\bbbz)}

\def\z0{{\bf z_0}}

\newcommand{\bit}{\begin{itemize}}
\newcommand{\eit}{\end{itemize}}
\newcommand{\no}{\noindent}

\begin{document}
\setcounter{page}{1}
\pagestyle{plain}

\section{The Puzzle In A Nutshell}

First things first. The puzzle is that an M5-brane wrapping a four-cycle holomorphically 
embedded in $\C^4$ is different from all other M-branes wrapped on holomorphic cycles \cite{Me}.  
This difference is manifest in several ways\footnote{It was pointed out in \cite{Julie} that an M5-brane
wrapping a four-cycle is inconsistent because of anomalies} and proves to be a special nuisance 
when we attempt to find the corresponding supergravity solution. In this note we will examine this 
unusual wrapped M5-brane, using the information encoded in its supergravity solution. However, 
before we attempt to resolve the puzzle of {\bf why} this brane is so different, we will first describe 
briefly the ways in which it stands out from the crowd. 

Often, a system of self-intersecting M$p$-branes corresponds to the singular limit of a single 
M$p$-brane wrapping a smooth cycle. More precisely speaking, this is the case 
when the cycle in question is described by factorisable embedding functions;  each  
factor then specifies the world-volume of one constituent brane.  For an M$p$-brane 
wrapped on a holomorphic cycle, the configuration obtained in the limit when the cycle becomes
singular is given by a number of planar M$p$-branes intersecting orthogonally along 
$(p-2)$ spatial directions, as expected by the $(p-2)$ self intersection rule \cite{p-2}, \cite{noforce}.

Rules however, do have exceptions and in the M5-brane configuration under study here, 
the $(p-2)$-rule has finally met its Waterloo. There is no way in which an M5-brane 
wrapped on a holomorphic four-cycle in $\C^4$ can be realised as a system of orthogonally 
intersecting M5-branes, with each pair  of branes intersecting in 3 (spatial) directions. 

At least one pair of fivebranes with a string intersection must be included in order for the resulting 
configuration to 'smoothen out' into a single fivebrane wrapped on a holomorphic four-cycle in
$\C^4$. The simplest intersecting brane realisation of this wrapped M5-brane is in fact given by a 
single pair of orthogonal fivebranes which share only one spatial direction, as shown below.

\be
\begin{array}[h]{|c|cc|cccc|cccc|c|}
  \hline
   \; & 0 & 1 & 2 & 3 & 
              4 & 5 & 6 & 7 & 8 & 9 & 10\\
  \hline
  {\bf M5} & \times & \times & \times & \times & \times & \times
               &  &  &  &  &  \\
  {\bf M5} & \times & \times 
               &  &  &  &  & \times & \times & \times & \times  & \\
  \hline
\end{array}
\label{mmaker}
\ee

But the anarchy does not end here. This brane configuration, not yet happy with the havoc it has 
caused, proceeds to also turn the harmonic function rule on its head! The harmonic function 
rule \cite{Tseytlin}, a prescription for constructing the supergravity solution for 
systems of intersecting branes,  gets its name from the fact that the 
contribution of each constituent brane to the resulting solution can be 
expressed in terms of a harmonic function. One of the hallmarks of this rule
was that the harmonic functions depend only on coordinates, 
which are simultaneously transverse to every brane in the system. 

Based on experience, we would thus expect the supergravity solution for the configuration 
(\ref{mmaker})  to depend only on $X^{10}$. We find however, that the solution is completely 
indepent of this overall transverse direction and the harmonic functions depend instead on the 
relative transverse directions ($X^2 \dots X^9$).  This unparalleled behaviour is in itself 
enough to indicate that there is something special going on with this system.

All these factors put together make for a rather complicated situation; one which is not at all well 
understood. It is in an attempt to de-mystify this intersecting brane system that we turn to its 
'smooth' version; an M5-brane wrapped on a holomorphic four-cycle in $\C^4$. We explore 
the corresponding supergravity solution, hoping to pick up clues which will help us uncover the 
underlying reason why this configuration breaks all the rules. 

\section{A Tool Kit}

Before we can expect to answer a question, we must first understand exactly what it is
that we are asking.  In formulating our question more precisely, we will also
get some valuable hints on where to begin looking for an answer. In this section, we will 
essentially collect tools which will help us in the process of both asking the question, and later, 
answering it. We first skim over some basic background which is relevant for the problem at hand.  

We start with a short review of planar BPS M-branes and the corresponding supergravity 
solutions \cite{LecNotes}. This allows us to set notation and also to remind the reader of some 
facts which will be used later. We then move on to the harmonic function rule, which enables us 
to build supergravity solutions for more complicated BPS brane configurations in M-Theory.  
Applying this rule to the system (\ref{mmaker}) we are lead to several contradictions.  By exploring 
these contradictions we gain a deeper understanding of the issues we need to address. 

\subsection{Understanding the Question}

Half  BPS planar M2-branes and M5-branes give rise to a large number of supersymmetric states in 
M-Theory. These states can be generated either by wrapping flat M-branes on supersymmetric 
cycles, or by building configurations of intersecting branes. 

\begin{center}
\underline{\bf \sf Supergravity Solutions for Planar M-Branes}:
\end{center}

\no
Before we present the actual solutions for flat M-branes, we pause for a moment to discuss the 
general features we expect to these solutions to contain. Since the world-volume of a planar M-brane 
respects Poincare invariance, the metric describing the surrounding space-time should be 
independent of coordinates tangent to the M-brane. Also, the configuration is invariant under 
rotations in 
the transverse directions; this is reflected in the fact that the metric is a function of only the 
radial coordinate $r$ in the transverse space. Under these conditions, the equation of motion for the 
field strength $d * F = 0$ reduces to the requirement that the function characterising the metric
is in fact a {\bf harmonic} function of $r$.  All these features are manifest in the solutions given 
below. 

Planar M5 and M2-branes are massive objects which are charged under the supergravity three-form. 
When placed in flat spacetime they deform it in such a way that the resulting background can be 
described as follows:

\bea
\begin{array}[h]{|cc|cc|}
\hline 
& & & \\
{\bf M5-brane} & & & ds^2 = H^{-1/3} \eta_{\mu \nu} dX^{\mu} dX^{\nu} +
H^{2/3} \delta_{\alpha \beta} dX^{\alpha} dX^{\beta} \\
& & & {\rm and} \; \; F_{\alpha \beta \gamma \delta} =
\frac{1}{2} \epsilon_{\alpha \beta \gamma \delta \rho}
\partial_{\rho} H \; \; 
{\rm where} \; H = 1 + \frac{a}{r^3}
\label{flatm5}\\
& & & \\
\hline
& & & \\
{\bf M2-brane} & & & ds^2 = H^{-2/3} \eta_{\mu \nu} dX^{\mu} dX^{\nu} +
H^{1/3} \delta_{\alpha \beta} dX^{\alpha} dX^{\beta} \\
& & & {\rm and} \; \; F_{{\mu}_0{\mu}_1{\mu}_2 \alpha} =
\frac{ \partial_{\alpha} H}{2 H^2} \; \; 
{\rm where} \; H = 1 + \frac{a}{r^6} \\
& & & \\
\hline
\end{array}
\eea
In the expressions above, $X^{\mu}$ denotes coordinates tangent to a brane and 
$r$ is the radial coordinate in the transverse space spanned by coordinates $X^{\alpha}$.

Planar M-branes, being half BPS objects, preserve 16 real
spacetime supersymmetries. These correspond to the components of a 
spinor $\chi$ which satisfies the condition 
${\hat \Gamma}_{{\mu}_0{\mu}_1{\mu}_2{\mu}_3{\mu}_4{\mu}_5} \chi = \chi$ 
for an M5-brane with worldvolume $X^{{\mu}_0} \dots X^{{\mu}_5}$, and 
${\hat \Gamma}_{{\mu}_0{\mu}_1{\mu}_2} \chi = \chi$ 
for a flat M2-brane spanning directions $X^{{\mu}_0}X^{{\mu}_1}
X^{{\mu}_2}$.\\ 

\newpage 

\begin{center}
\underline{\bf \sf Harmonic Function Rule}
\end{center}

\no
A system of M-branes gives rise to a bosonic background which combines the individual effects
of each constituent brane. If the branes are aligned in a certain way, the resulting background 
can still be a solution to the equations of motion of  D=11 supergravity.  Since the  solution for a 
single brane is expressible in terms of a harmonic function, an extremal (no binding energy) 
BPS configuration of N branes is expected to have a supergravity solution characterised by 
N independent harmonic functions.  The harmonic function rule \cite{Tseytlin} gives us a way in 
which to obtain the supergravity solution of an intersecting brane system by superposing the 
individual bosonic fields arising from each component brane. 

Before we specify the rule, here is some notation: In the presence of an intersecting M-brane
configuration,  spacetime can naturally be 'divided' into three subspaces.  The directions common 
to all constituent M-branes live in the common tangent subspace.  The relative transverse subspace 
is spanned by directions which are tangent to at least one but not all of the planar M-branes (From 
the wrapped brane point of view, this is known as the embedding space, as it is where the 
supersymmetric cycle wrapped by the M-brane is embedded).  Finally, there is the overall 
transverse subspace which spans directions transverse to all constituent branes in the intersecting 
brane system.\\

\no
{\bf The Field Strength}:
The field strength components due to each constituent M-brane carry different indices, hence the 
field strength of the resulting intersecting brane configuration can be obtained merely by adding 
the individual field strengths corresponding to each M-brane.\\

\no
{\bf The Metric}:
Assigning a harmonic function to each constituent M-brane, the metric for an orthogonally 
intersecting M-brane system can be written in the form
\be
ds^2 = \sum_I f_I (X^{\alpha}) \; \eta_{IJ} dX^I dX^J
\ee
where $I, J$ run over all spacetime and $\alpha$ is used to label the overall
transverse directions. In analogy to the single brane case, the coefficient $f_I$ contains a factor 
$H^{-1/3}$
due to each M5-brane  (and  $H^{-2/3}$ due to each M2-brane) whose world-volume includes 
the coordinate $X_I$.  Similarly, every M5-brane lying transverse to $X_I$ contributes a factor of 
$H^{2/3}$ (whereas every transverse M2 brane contributes $H^{1/3}$) to $f_I$. 

\begin{center}
\underline{\bf \sf An Example}:\\
\end{center}

\no
In order to illustrate the application of the harmonic function rule, 
consider the following system of two M5-branes intersecting in three spatial directions.
$$
\begin{array}[h]{|c|cccc|cccc|ccc|}
  \hline
   \; & 0 & 1 & 2 & 3 & 
              4 & 5 & 6 & 7 & 8 & 9 & 10\\
  \hline
  {\bf M5} & \times & \times & \times & \times & \times & \times
               &  &  &  &  &  \\
  {\bf M5} & \times & \times & \times & \times
               &  &  & \times & \times &  &  &  \\
  \hline
\end{array}$$
The harmonic function rule then dictates the following metric
\bea
ds^2 &=& H_1^{-1/3}  H_2^{-1/3} 
(- dX_{0}^2 + dX_{1}^2 + dX_{2}^2 + dX_{3}^2) + H_1^{-1/3} H_2^{2/3} (dX_{4}^2 + dX_{5}^2) 
\nonumber \\ &+& 
H_1^{2/3} H_2^{-1/3}  (dX_{6}^2 + dX_{7}^2) 
+ H_1^{2/3}  H_2^{2/3} (dX_{8}^2 + dX_{9}^2 +dX_{10}^2) \nonumber
\eea
and field strength components
$$ F_{67 \alpha \beta} = \epsilon_{\alpha \beta \gamma} H_1 \;\;\;\;
F_{45 \alpha \beta} = \epsilon_{\alpha \beta \gamma} H_2$$
where the functions $H_1$ and $H_2$ depend only on $X^{\alpha}$ for 
$\alpha = 8,9,10.$\\ 

\subsection{Spelling out the Problem}

Applying the harmonic function rule to the system (\ref{mmaker}) leads to the
 metric:
\bea
ds^2 &=&  
H_1^{-1/3}  H_2^{-1/3} (- dX_{0}^2 + dX_{1}^2) 
+  H_1^{-1/3}  H_2^{2/3} (dX_{2}^2 + dX_{3}^2 + dX_{4}^2 + dX_{5}^2)
\nonumber \\
&+&  H_2^{1/3} H_2^{-1/3} (dX_{6}^2 + dX_{7}^2 + dX_{8}^2 + dX_{9}^2) 
+ H_1^{2/3}  H_2^{2/3} dX_{10}^2
\label{mmakermetric}
\eea
The scarcity of overall transverse directions leads to several glaring problems. The most 
immediate issue is of course that if $H_1$ and $H_2$ are functions of $X^{10}$ alone, then
 they cannot possibly be harmonic functions and still have the required fall-off at infinity.  
Moreover,  it is completely unclear what the expressions for the field strength components 
should be;  we obviously cannot proceed in analogy to the example worked out above.  

A rather unusual solution was proposed in \cite{monster}.  The form of the metric 
(\ref{mmakermetric}) was left unmodified, but the roles of the overall transverse and relative 
transverse directions were interchanged.  Rather than depending on the overall tranverse 
direction, it was suggested that the functions in the metric anstaz depend instead on the relative 
transverse directions. In particular, $H_1$ was  found to be a harmonic function of  
$X^6 \dots X^9$, and $H_2$ was required to solve the flat space Laplacian in $X^2 \dots X^5$.  
Imposing this functional dependence did make it possible to compute a supergravity solution, 
but it did not shed any light on why, in this particular case,  the harmonic functions behave in 
such a  strange manner.  This is the question which served as the motivation for 
the work presented in this note, and hence it is a question we will return to later. 

\subsection{Looking for an Answer}

Having now made the puzzle explicit, we begin our search for the resolution. We start by adding to 
our tool-kit of concepts a few basic ideas which will save us a lot of labour in our quest for an answer, 
and also provide us with the necessary vocabulary in which to formulate the results of our search. 

\begin{center}
\underline{\bf \sf Spinors on Complex Manifolds}
\end{center}

\no
Spinors on complex manifolds can be expressed as Fock space states, using the fact that the 
Clifford algebra in flat compex space resembles the algebra of fermionic creation and annihilation 
operators. More explicitly, the Clifford algebra in $\C^n$ takes the form
\be
\{ \Gamma_{z_i}, \Gamma_{{\overline{z}}_j} \} = 2 \eta_{i \bar{j}}.
\ee
where $i, j = 1 \dots n$ and we have defined the 
$\Gamma$ matrices for a complex coordinate $z_j = x_j + iy_j$ as follows:
\bea
\Gamma_{z_j} &=& \frac{1}{2} (\Gamma_{x_j} + i \Gamma_{y_j})\\ \nonumber
\Gamma_{{\overline{z}}_j} &=& \frac{1}{2} (\Gamma_{x_j} - i \Gamma_{y_j})
\eea
Declaring $\Gamma_{z_i}$ to be creation operators and $\Gamma_{{\overline{z}}_j}$ 
to be annihilation operators, a Fock space can be generated by acting the creation 
operators on a vacuum. Because there are $n$ creation operators, each state in the
Fock space is labelled by $n$ integers taking values 0 or 1 which correspond to its 
fermionic occupation numbers. It will later prove very useful to express 
Killing Spinors in this way. 

\begin{center}
\underline{\bf \sf The Fayyazuddin-Smith Ansatz}
\end{center}

\no
By its very construction, the harmonic function rule does not extend to supergravity 
solutions for branes localised in the relative transverse directions. In an attempt to find 
supergravity solutions for such localised brane intersections Fayyazuddin and Smith 
\cite{FS} proposed a metric ansatz by applying symmetry based arguments to the 
background of a curved M-brane.

When an M$p$-brane wraps a supersymmetric $m$-cycle, part of the world-volume of the 
brane will in general remain unwrapped.  Poincare invariance is expected to hold along 
these $(p + 1 - m)$ unwrapped (common tangent) directions $X^{\mu}$,
implying that the metric should be independent of these coordinates.  Rotational 
invariance in the overall transverse directions $X_{\alpha}$ leads to a diagonal metric  
in this subspace and further dictates that the undetermined functions in the metric ansatz 
depend only on  $\rho = \sqrt{X_{\alpha} X^{\alpha}}$. A complex structure can be defined 
in the remaining (relative transverse) directions. The Hermitean metric $G_{M {\bar N}}$ 
in this subspace cannot be constrained using the isometries of the brane configuration.  

A metric incorporating all the above features takes the form
\be
ds^2 = H_1^2  \eta_{\mu \nu }dX^{\mu} dX^{\nu} + 2 G_{M {\bar N}} dZ^{M}
dZ^{\bar N} + H_2^2 \delta_{\alpha \beta} dX^{\alpha} dX^{\beta}
\ee
and has come to be known as the Fayyazuddin-Smith metric ansatz. 

\begin{center}
\underline{\bf \sf Looking for Supergravity Solutions}
\end{center}

\no
The fact that a configuration preserves supersymmetry implies that 
the supersymmetric variation $\delta_{\chi} \Phi$  vanishes for all fields $\Phi$ 
when the variation parameter $\chi$ is a Killing spinor of the background. 

Denoting flat indices in 11-dimensional spacetime by $i,j$ and curved indices by $I,J$ the 
bosonic part of the action for 11d supergravity can be written as
\be
S = \int d^{11}X {\sqrt {- G}} \{
R - \frac{1}{12} F^2 - \frac{1}{432} \epsilon^{I_1....I_{11}}
F_{I_1...I_4} F_{I_5....I_8} A_{I_9..I_{11}}
\}
\ee
Throughout this note we will be dealing solely with bosonic backgrounds,
fermionic fields have been set to zero and the supersymmetric variations of the 
metric and four-form field strength vanish identically. The only non-trivial 
requirements  for supersymmetry preservation arise from the gravitino
variation equation
\be
\delta\Psi_{I} = (\partial_{I}  + \frac{1}{4} \omega_I^{ij} \hat{\Gamma}_{ij}
+ \frac{1}{144}{\Gamma_{I}}^{JKLM}F_{JKLM}
-\frac{1}{18}\Gamma^{JKL}F_{IJKL})\chi. \label{susy}
\ee
	
We could also turn the logic around. By requiring $\delta_{\chi} \Psi = 0$
to hold for a given metric, we find a set of relations between  the metric and  
field strength of the supergravity three-form which must be true if supersymmetry is 
to be preserved by the background.   If the four-form obeying these relations is such that 
$dF=0$ and $d \star F =0$, then Einstein's equations are guaranteed to be satisfied and 
we have determined the bosonic components of a BPS solution to 11-dimensional supergravity.

This is the procedure followed in the proceeding section, to analyse an M5-brane 
wrapping a holomorphic 4-cycle in $\C^4$.   We enforce supersymmetry preservation 
for a metric of the form proposed by Fayyazuddin and Smith. Expressing $\chi$ as a 
sum of Fock space states,  $\delta \Psi_{I} = 0$ reduces to a sum of linearly independent 
relations each of which must be put to zero seperately. Satisfying these relations is the 
first step to finding the supergravity solution corresponding to the M-brane in question. 

\section{The Mischief Maker}

Having a complete tool kit at our disposal, we are now in a position to break down our 
problem and study its various components.  We begin by writing down the 
Fayyazuddin-Smith ansatz for the metric describing the supergravity background 
created by an M5-brane wrapping a holomorphic 4-cycle in $\C^4$. As can be seen from 
the discussion in the previous section, the relevant metric ansatz is 
\be
ds^2 = H_1^2  \eta_{\mu \nu }dX^{\mu} dX^{\nu} + 2 G_{M {\bar N}} dZ^{M}
dZ^{\bar N} + H_2^2 dy^{2}
\label{standard}
\ee
where $Z^{M}$ denote the holomorphic coordinates $s, u, v, w$ in $\C^4$ and $y$ is the overall
transverse direction. In order for Lorentz invariance to be preserved along the unwrapped directions 
of the brane's worldvolume, the functions $H_1, H_2$ and 
$G_{M {\bar N}}$ must be independent of $X_{\mu} = X^0$ and $ X^1$. 

\subsection{Killing Spinors}

The amount of supersymmetry preserved by a $p$-brane with 
worldvolume $X^{M_0}X^{M_1}...X^{M_p}$ is given by the number of spinors which satisfy the 
equation \cite{BBS}
\be
\chi = \frac{1}{(p+1)!} \epsilon^{\alpha_0\alpha_1 ... \alpha_p}
\Gamma_{M_0M_1 ... M_p} 
\partial_{\alpha_0} X^{M_0}\partial_{\alpha_1} X^{M_1} ....
\partial_{\alpha_p} X^{M_p} \chi
\label{KSeqn}
\ee
where $\Gamma_{M_0M_1 ... M_p}$ is the completely anti-symmetrized 
product of $p+1$ eleven dimensional $\Gamma$ matrices. 

Consider now the metric (\ref{standard}).  An 11-dimensional spinor in this background can 
be decomposed into spinors within and transverse to $\C^4$. The spinors on $\C^4$ 
can then be realised as linear combinations of Fock space states, as explained earlier. Using this 
construction,  Killing spinors $\chi$ of the spacetime (\ref{standard}) can be expressed as a sum 
of terms of the form $\alpha \otimes \psi \equiv \alpha \otimes |n_s n_u, n_v, n_w>$ where 
$\alpha$ is a three-dimensional spinor and $n_z  ({\rm for} \; z=s, u,v,w)$ are the fermionic 
occupation numbers of the state, corresponding to the action of $\Gamma_{z}$ on the Fock vacuum.

From (\ref{KSeqn}) we see that the Killing spinors for an 
M5-brane wrapping a supersymmetric four-cycle in $\C^4$ are such that \cite{wrapM5}
\be
\epsilon^{abcd}
\Gamma_{0 1} \Gamma_{m \bar{n}  p \bar{q}}
\partial_{a} X^{m} \partial_{b} X^{\bar{n}}
\partial_{c} X^{p} \partial_{d} X^{\bar{q}} \chi =
\chi
\ee
where the $\Gamma_m$ are flat space $\Gamma$-matrices and ${\sigma}^a \dots {\sigma}^d$ 
are coordinates on the four-cycle. This can be simplified and expressed as follows
\be
\Gamma_{01}\Gamma_{ m \bar{n} p \bar{q}} \chi =
(\eta_{m \bar{n}} \eta_{ p \bar{q}} -
\eta_{m \bar{q}} \eta_{p \bar{n}} )
\chi
\label{killspinors}
\ee
where $\eta_{m \bar{n}}$ is the flat space metric in $\C^4$. 
A solution to this equation is given by the Majorana spinor $\chi$ such that
\be
\chi = \alpha \otimes (|0000> + |1111>)
\label{chi}
\ee
where the chirality of the spinor $\alpha$ in the space-time
transverse to $\C^4$ is fixed by the requirement that
\be
\Gamma_{01} \; \alpha = \alpha \\
\label{chirality}
\ee

In order to count the number of supercharges preserved by this configuration, note that 
as a generic spinor $\chi$ would have had 32 complex components: 2 complex components 
come from the Dirac spinor $\alpha$ and there are 16 linearly independent 
Fock space states in $\C^4$ . After the constraints (\ref{killspinors}) are imposed, only 
2 of these 16 components survive and $\chi$ is left with 4 complex degrees of freedom. 
Determining the chirality of $\alpha$  cuts the degrees of freedom down by half and 
enforcing the Majorana condition cuts them down by a further half.  The wrapped 
M5-brane thus preserves $1/16$ of the spacetime supercharges, corresponding to the 
2 spinors which satisfy the above conditions.\\

\subsection{Supersymmetry Preservation Conditions}

Contrary to the way things are normally done, we will consider for the moment 
that all possible components of the four-form field strength could in principle be turned 
on. Conventional wisdom dictates that since fivebranes couple to the supergravity 
three-form purely magnetically, only magnetic components of the field strength (i.e, 
those with indices purely transverse to the worldvolume) will be present. However, 
conventional wisdom hasn't exactly served us very well  when it comes to this 
particular configuration, so for now we keep the electric components in the picture.  
The only arguments we admit at this point are those of symmetry. 
Components like $F_{\mu A B C}$ (where $A,B,C$ take values in $C^4$ and $y$) 
would destroy the $SO(1,1)$ isometry expected of the solution and are thus set to zero.  

We proceed to look for BPS solutions to $d = 11$ supergravity by demanding that the 
supersymmetry variation of the gravitino vanishes for the metric ansatz (\ref{standard}) 
and Killing spinor (\ref{chi}). This gives us a set of equations which can be solved to obtain
expressions for the field strength components. In addition,  we also find a constraint on the 
metric and a relation between the metric and the Killing spinor. These are given below. \\

\no
\underline{\bf \sf Metric Constraint}\\

\no
The metric is subject to the constraint
\be
\partial [G^{-\frac{2}{3}} {H_1}^{-\frac{10}{3}} H_2 \omega \wedge \omega
\wedge \omega ] = 0.
\ee
Here, $\sqrt{G}$ denotes the  determinant of the Hermitean metric in the complex subspace, and 
$\omega = i G_{M {\bar N}} dZ^M dZ^{\bar N}$ is the associated two-form. \\

\no
\underline{\bf \sf Killing Spinors}\\

\no
The Killing spinor $\chi$ is specified through (\ref{chi}) once we 
determine $\alpha$. We already know that $\alpha$ is proprotional to a constant
spinor whose chirality is fixed through (\ref{chirality}).  All that needs to be found in order to 
uniquely specify $\alpha$ is the factor multiplying the constant spinor. Supersymmetry 
preservation dictates this factor be such that
\be
 24 \partial_{\bar R} ln \alpha = 
\partial_{\bar R} ln G + 20 \partial_{\bar R} ln H_1
\ee

\no
\underline{\bf \sf Field Strength Components}\\

The gravitino variation equations set the components $F_{01MN}$ and 
$F_{M \bar{N} \bar{P} \bar{Q}}$ to zero identically. The remaining field strength 
components can roughly be catagorized as follows. To begin with, there are components 
with indices lying along the world-volume of the brane; these will be referred 
to in the following as electric type components, since they imply that the wrapped 
M5-brane also couples {\bf electrically}  to the supergravity three-form, defying intuition. 
These components are:
\be
F_{01 {\bar R} y} = \frac{1}{2} \partial_{\bar R} 
({H_1}^2 H_2) 
\ee
and 
\be
F_{01 M {\bar N}} = \frac{H_1}{2 H_2} \partial_{y} (H_1 G_{M {\bar N}}) 
\ee\\

\no
Then there are components with a more familiar structure. These have 
indices in the (relative and overall) transverse  directions only, indicating that the fivebrane 
couples magnetically to the supergravity three-form, as expected. By solving the gravitino variation 
equations, we are able to determine some of these components completely and to impose contraints on 
the others. Two components for which we are able to obtain  explicit expressions are
\be
F_{suvw} = - \frac{\sqrt{G}}{4 H_2} [\partial_{y} ln G + 
8 \partial_{y} ln H_1]
\ee
and 
\be
F_{M P {\bar Q} {\bar N}} = \frac{1}{6 H_2 H_1^{5}}
\partial_y  [H_1^{5} G_{P {\bar Q}} G_{M {\bar N}} - 
G_{M {\bar Q}} G_{P {\bar N}})] 
\ee\\

\no
The component $F_{M{\bar N} {\bar P} y}$ is determined in terms of $F_{MNPy}$ by the following 
equation
\bea
 G_{M {\bar Q}}[9 \partial_{\bar P} ln H_2 - 
\partial_{\bar P} ln G - 8 \partial_{\bar P} ln H_1]
- G_{M {\bar P}}[9 \partial_{\bar Q} ln H_2 - 
\partial_{\bar Q} ln G - 8 \partial_{\bar Q} ln H_1]  \nonumber \\
= 9 [\partial_{\bar Q} G_{M {\bar P}} - \partial_{\bar P} 
G_{M {\bar Q}}] 
- 6 H_2^{-1} {\epsilon}^{RN}_{\; \; \; \; \; \; {\bar P} 
{\bar Q}} F_{MRNy}
- 18 H_2^{-1}  F_{M {{\bar P} {\bar Q}} y}
\label{Mnpy}
\eea\\

\no
The only hitch is that $F_{MNPy}$ cannot be fixed yet. However, we know that it is subject to 
\be
H_2^{-1} {\epsilon}^{MNP}_{\; \; \; \; \; \; \; \; \; {\bar R}} 
F_{MNPy} - 
\partial_{\bar R} ln G - 8 \partial_{\bar R} ln H_1 = 0
\label{MNPy}
\ee
Once this constraint is solved and $F_{MNPy}$ is known, $F_{M{\bar N} {\bar P} y}$ can be obtained 
immediately from (\ref{Mnpy}) and the four-form will then be completely determined. 

\subsection{Solutions to $\delta \Psi = 0$}
\label{solutions}

It is not a trivial task to solve the constraint $(\ref{MNPy})$ and thereby obtain the most general 
expressions for all field strength components.  From now on we will restrict ourselves to a subcase, 
perhaps the simplest possible one,  in which  $(\ref{MNPy})$ is satisfied by setting $F_{MNPy} = 0$.  
While doing so, we must keep in mind that the expressions we obtain follow only from the 
requirement of supersymmetry preservation; they cannot be said to comprise a supergravity 
solution until we impose the equations of motion $d F = 0$ and check that Bianchi identity 
$d * F + F \wedge F = 0$ is also satisfied. \\

\no
\underline{\bf \sf Assumptions}\\

\no
The vanishing of $F_{MNPy}$ implies that
\bigskip
\begin{center}
\noindent\fbox {\noindent\parbox{3.7in}
{$$ \partial_{\bar R} ln G + 8 \partial_{\bar R} ln H_1 = 0$$}}
\end{center}
\bigskip
This can be then substituted into the expressions found in the 
previous section in order to yield the set of equations given below\\

\no
\underline{\bf \sf Metric Constraint}\\

\no
The Hermitean metric is subject to the constraint
\bigskip
\begin{center}
\noindent\fbox {\noindent\parbox{3.7in}
{$$
\partial [H_1^{2} H_2 \; \omega \wedge \omega
\wedge \omega ] = 0$$}}
\end{center}
\bigskip

\no
\underline{\bf \sf Field Strength Components}\\

\no
The four-form is specified by the following components
\bigskip
\begin{center}
\noindent\fbox {\noindent\parbox{3.7in}
{$$
F_{01 {\bar R} y} = \frac{1}{2} \partial_{\bar R} 
({H_1}^2 H_2)$$

$$F_{01 M {\bar N}} = \frac{H_1}{2 H_2} \partial_{y} (H_1 G_{M {\bar N}})$$

$$F_{M {{\bar P} {\bar Q}} y} = 
\frac{1}{2} [\partial_{\bar Q} (H_2 G_{M {\bar P}}) - \partial_{\bar P} 
(H_2 G_{M {\bar Q}})] $$

$$ F_{M N {\bar P} {\bar Q}} = \frac{1}{6 H_2 H_1^{5}}
\partial_y  [H_1^{5}  ( G_{P {\bar N}} G_{M {\bar Q}} - 
G_{M {\bar N}} G_{P {\bar Q}})] $$}}

\end{center}
\bigskip
All other components vanish.\\

\no
\underline{\bf \sf Killing Spinors}\\

\no
The Killing spinor $\chi$ can be obtained from (\ref{chi}) using
$$2 \partial_{\bar R} ln \alpha =  \partial_{\bar R} ln H_1$$

\section{A Consistency Check}

If this work is to be considered a step forward, it should contain the information we already 
knew and also expand upon our previous knowledge. At the very least then, we should be 
able to recover  the supergravity solution constructed via the harmonic function rule. Before we 
can show that this is so, we must first remind ourselves of what the harmonic function rule 
said in the first place.

\newpage

\subsection{What We Knew}

\begin{center}
\underline{\bf M5 $\perp$ M5 (1)}
\end{center}

\no
Consider an brane configuration made up of two flat M5-branes which intersect in a string. 
Let them have worldvolumes $01s\bar{s}u\bar{u}$ and $01v\bar{v}w\bar{w}$
respectively, where $s,u,v,w$ are the holomorphic coordinates spanning $\C^4$. 
If the harmonic functions corresponding to
the two branes are denoted by $H_A$ and $H_B$, the harmonic 
function rule dictates that the metric should take the form \cite{monster}

\bea
ds^2 = {H_A}^{-\frac{1}{3}} {H_B}^{-\frac{1}{3}} (-dX_0^2 + dX_1^2) +  
{H_A}^{-\frac{1}{3}} {H_B}^{\frac{2}{3}} (ds d\bar{s} + du d\bar{u}) 
\nonumber \\ 
{H_A}^{\frac{2}{3}} {H_B}^{-\frac{1}{3}} (dv d\bar{v} + dw d\bar{w}) 
+ {H_A}^{\frac{2}{3}} {H_B}^{\frac{2}{3}} dy^{2}\;\;\;\;\;\;\;\;\;\;\;
\label{metricm5m5}
\eea
However, there are a few surprises in store. It turns out that both
harmonic functions are independent of the overall transverse 
direction $y$, in contradiction to the normal mode of 
operation of the rule which generally leads the harmonic functions to 
depend {\bf only} on the overall transverse coordinates! What happens 
in this case however, is that $H_A$ is a function of 
$v, w, {\bar v}, {\bar w}$ and $H_B$ of $s, u, {\bar s}, {\bar u}$. Being
harmonic functions, they are solutions to the corresponding flat space
Laplace equations:
\bea
(\partial_v \partial_{\bar v} + \partial_w \partial_{\bar w}) H_A = 0 \nonumber
\\
(\partial_s \partial_{\bar s} + \partial_u \partial_{\bar u}) H_B = 0
\eea
The non-vanishing components of the field strength are
\bea
F_{s {{\bar s} {\bar u}} y} = 
\frac{1}{4} \partial_{\bar u} H_B \;\;\;\;\;
F_{u {{\bar u} {\bar s}} y} = 
\frac{1}{4} \partial_{\bar s} H_B \nonumber \\
F_{v {{\bar v} {\bar w}} y} = 
\frac{1}{4} \partial_{\bar w} H_A \;\;\;\;\; 
F_{w {{\bar w} {\bar v}} y} = 
\frac{1}{4} \partial_{\bar v} H_A 
\label{fstm5m5}
\eea
Though it remains a mystery why the harmonic function rule metric decides to localize the 
intersection in the relative transverse directions, it was suggested in \cite{TheG} that
perhaps one of the curious features of this solution could be explained.
The proposal was that the lack of  dependence on the overall transverse direction signals 
the presence of a membrane.  This membrane was conjectured to stretch between the 
two fivebranes, intersecting each in a string. The fivebranes are then brought closer and 
closer together until the membrane collapses and all that is left of it is the string intersection
on the fivebranes; this intersection, it was said, should be interpreted as a collapsed membrane. 

While this is definitely a possibility, it is probably not the sole possibility. However, we 
will argue that other options exist later. First let us see how the supergravity solution for 
the system is modified upon the inclusion of a membrane.\\

\newpage

\begin{center}
\underline{\bf M5 $\perp$ M5 $\perp$ M2 (1)}
\end{center}

\no
Since the two fivebranes are exactly the same as in the previous 
configuration and the only addition is that of a membrane with
worldvolume $01y$, the corresponding metric is easily written down. 
If the membrane is characterized by a harmonic function $H_C$, 
the metric is \cite{monster}
\bea
ds^2 = {H_A}^{-\frac{1}{3}} {H_B}^{-\frac{1}{3}} {H_C}^{-\frac{2}{3}} 
(-dX_0^2 + dX_1^2) +  {H_A}^{-\frac{1}{3}} {H_B}^{\frac{2}{3}} 
{H_C}^{\frac{1}{3}}
(ds d\bar{s} + du d\bar{u}) \nonumber \\ 
{H_A}^{\frac{2}{3}} {H_B}^{-\frac{1}{3}} {H_C}^{\frac{1}{3}}
(dv d\bar{v} + dw d\bar{w}) 
+ {H_A}^{\frac{2}{3}} {H_B}^{\frac{2}{3}} {H_C}^{-\frac{1}{3}} dy^{2} 
\;\;\;\;\;\;\;\;\;
\label{metricm5m5m2}
\eea
Once more, we find $H_A = H_A(v, w, {\bar v}, {\bar w})$ and 
$H_B = H_B(s, u, {\bar s}, {\bar u})$. $H_C$ on the other hand,
while still independent of $y$, can depend on all the coordinates in 
the complex space.

In addition to the components (\ref{fstm5m5}) which are of course still 
present, the four-form field strength has two new components, given by
\be
F_{01yM} = \frac{1}{2} \frac{\partial_I H_C}{H_C^2}
\label{fstm5m5m2}
\ee
(where $M$ takes values in $\C^4$) and its complex conjugate. 
While $H_A$ and $H_B$ are conventional harmonic functions, in that they 
obey the flat space Laplace equations, $H_C$ is instead a generalised 
harmonic function since it obeys the following differential equation
\be
[H_A (\partial_s \partial_{\bar s} + \partial_u \partial_{\bar u})
+ H_B (\partial_v \partial_{\bar v} + \partial_w \partial_{\bar w})] H_C = 0
\label{cshfn}
\ee
It can now be seen that (\ref{metricm5m5}) and (\ref{fstm5m5}) 
are contained within (\ref{metricm5m5m2}) and (\ref{fstm5m5m2}) and 
may be obtained simply by putting $H_3 = 1$ in the latter equations. 

\subsection{What We Know Now}

In order to build up our faith in the analysis of Section 3, we check for its consistency with the 
results of the harmonic function rule reviewed above. 
One way of doing this is to study what emerges from the equations of Section 3.3 when we 
plug in the metric (\ref{metricm5m5m2}). Comparing this to the standard form 
$$ ds^2 = H_1^2  dX_{\mu} dX_{\nu} + 2 G_{M {\bar N}} dZ^{M}
dZ^{\bar N} + H_2^2 dy^{2}$$ of the Fayyazuddin-Smith ansatz, we find
\bea
H_1 = (H_A H_B {H_C}^2 )^{-\frac{1}{6}} \;\;\;\;\;\;\;\;\;\;\;\;\;\;\;\;\;
H_2 = ({H_A}^2 {H_B}^2 {H_C}^{-1})^{\frac{1}{6}} 
\;\;\;\;\;\;\;\;\;\;\;\;\;\;\; \nonumber \\
2 G_{s \bar{s}} = 2 G_{u \bar{u}} = {H_A}^{-\frac{1}{3}} {H_B}^{\frac{2}{3}} 
{H_C}^{\frac{1}{3}}
\;\;\;\;\;\;\;\;
2 G_{v \bar{v}} = 2 G_{w \bar{w}} = {H_A}^{\frac{2}{3}} {H_B}^{-\frac{1}{3}} 
{H_C}^{\frac{1}{3}}
\eea
These relations can then be substituted in the expressions found in Section 
\ref{solutions} to obtain the following:\\ 

\newpage

\no
\underline{\bf \sf Metric Constraint}\\

\no
The metric constraint now implies that
\bea
\partial_u ( H_C^{-1}  G_{s {\bar s}} G_{v {\bar v}} G_{w {\bar w}}) = 0
&\Rightarrow &
\partial_u H_A = 0 \\
\partial_s ( H_C^{-1} G_{u {\bar u}} G_{v {\bar v}} G_{w {\bar w}}) = 0
&\Rightarrow &
\partial_s H_A = 0 \\
\partial_v (H_C^{-1} G_{s {\bar s}} G_{u {\bar u}} G_{w {\bar w}}) = 0
&\Rightarrow &
\partial_v H_B = 0 \\
\partial_w (H_C^{-1} G_{s {\bar s}} G_{u {\bar u}} G_{v {\bar v}}) = 0
&\Rightarrow &
\partial_w H_B = 0
\eea
So $H_A$ can be a function of only $v, w, {\bar v}, {\bar w}$ 
and $H_B$ can depend only on $s, u, {\bar s}, {\bar u}$\\

\no
\underline{\bf \sf Field Strength Components: Electric}\\

\no
The four-form field strength has components of the type
$F_{01 {\bar P} y}$, given by
\be
F_{01My} = - \frac{1}{2} \frac{\partial_M H_C}{H_C^2}
\label{ffirst}
\ee
\be
F_{01{\bar N}y} = - \frac{1}{2} \frac{\partial_{\bar N} H_C}{H_C^2}
\ee
The only other components with indices $01$ have the structure
$F_{01 M {\bar N}}$, as follows
\be
F_{01s\bar{s}} = F_{01u\bar{u}} = \frac{1}{2 (H_A H_B )^{1/2}}
\partial_y \left(\frac{{H_B}^{1/2}}{{H_A}^{1/2}} \right) 
\label{fss}
\ee
\be
F_{01v\bar{v}} = F_{01w\bar{w}} = \frac{1}{2 (H_A H_B )^{1/2}}
\partial_y \left(\frac{{H_A}^{1/2}}{{H_B}^{1/2}} \right) 
\label{fvv}
\ee\\

\no
\underline{\bf \sf Field Strength Components: Magnetic}\\

\no
The contributions of the type $F_{M {{\bar P} {\bar Q}} y}$, taking 
into account the functional dependence of $H_A$ and $H_B$ on 
the complex coordinates, are
\bea
F_{s {{\bar s} {\bar u}} y} &=& 
\frac{1}{2} [\partial_{\bar u} (H_2 G_{s {\bar s}})] = 
\frac{1}{4} \partial_{\bar u} H_B \\
F_{u {{\bar u} {\bar s}} y} &=& 
\frac{1}{2} [\partial_{\bar s} (H_2 G_{u {\bar u}})] = 
\frac{1}{4} \partial_{\bar s} H_B \\
F_{v {{\bar v} {\bar w}} y} &=& 
\frac{1}{2} [\partial_{\bar w} (H_2 G_{v {\bar v}})] = 
\frac{1}{4} \partial_{\bar w} H_A \\
F_{w {{\bar w} {\bar v}} y} &=& 
\frac{1}{2} [\partial_{\bar v} (H_2 G_{w {\bar w}})] = 
\frac{1}{4} \partial_{\bar v} H_A
\eea
And lastly, there are components of the form $F_{M N {\bar P} {\bar Q}}$. 
These are given below
\bea
F_{u s {\bar s} {\bar u} } &=& 
\frac{1}{24} H_A^{1/2} H_B^{1/2} H_C^{2} \; [
\partial_y\left( H_B^{1/2} H_A^{-3/2} H_C^{-1} \right)]\\
F_{v w {\bar w} {\bar v} } &=& 
\frac{1}{24} H_A^{1/2} H_B^{1/2} H_C^{2} \;
[\partial_y\left(H_A^{1/2} H_B^{-3/2} H_C^{-1} \right)]\\
F_{v s {\bar s} {\bar v} } &=& 
\frac{1}{24} H_A^{1/2} H_B^{1/2} H_C^{2} \;
[\partial_y\left( H_A^{-1/2} H_B^{-1/2} H_C^{-1} \right)] \label{flast}\\
&=& F_{w s {\bar s} {\bar w} } = F_{v u {\bar u} {\bar v} } 
= F_{w u {\bar u} {\bar w} } \nonumber
\eea

\subsection{And How It Fits Together}

It is obvious from the above analysis that by requiring supersymmetry to be preserved
in the presence of the metric (\ref{metricm5m5m2}), we end up with  more field strength components 
than were predicted by the harmonic function rule. Now that we have all 
the expressions infront of us, a possible solution to this conundrum appears. 

The harmonic function rule implicitly assumes that the only non-vanishing field 
strength components are those which arise from gauge potentials coupling ${\bf electrically}$ 
to each of the branes in the system. Since the metric (\ref{metricm5m5m2}) was supposed to
 describe two fivebranes and a membrane, the only components of the four-form field strength 
which were expected to be present were
$F_{M N {\bar P} {\bar Q}}, F_{M {{\bar P} {\bar Q}} y}$, $F_{01 {\bar P} y}$ and their 
complex conjugates. In particular, there was
no allowance for terms like $F_{01M {\bar N} }$.  As we have seen above, such terms {\bf do} in fact
arise and will not vanish, unless this condition is explicitly imposed. 

From the expressions above,  it is clear that $F_{01M {\bar N} } = 0$ can only be enforced by 
imposing $\partial_y H_A = \partial_y H_B = 0$. 
Plugging this restriction into the remaining equations, we find that the field strength 
components reduce to (\ref{fstm5m5m2}) and 
(\ref{fstm5m5}). The Bianchi Identity for the four-form implies that
$\partial_y H_C = 0$ and in addition requires $H_C$ to obey the curved space Laplace 
equation (\ref{cshfn}).  Hence, by requiring $F_{01M {\bar N} }$ to vanish,  
we are able to reproduce the expressions found earlier using the harmonic function rule.

\section{Justifications}

Given the discussion of the previous section, the origin of the $y$-independence seen
 in the harmonic function rule now becomes clear. By insisting that the four-form field strength 
can only have purely magnetic components for each constituent M5-brane, we 
are unnecessarily restricting ourselves to only a subclass of solutions; those which
are smeared in the overall transverse direction. Our intuition which dictates that 
M5-branes couple to the supergravity three-form purely magnetically 
has held true for all the systems we have encountered so far, but apparently it breaks down here. 

The reason for this is in fact quite simple. 
For all the M-brane systems we had studied up to now,  $F \wedge F$ was always zero and 
this is no longer true. The  strange behaviour of the harmonic function rule can also be 
explained when we recall that this rule was meant to be applied \cite{Tseytlin}  only to 
configurations 
where there were no Chern-Simmon terms in the equations of motion for the four-form. 
It is clear that the non-zero contribution of  $F \wedge F $  has far reaching consequences
 and is hence worth discussing in a bit more detail. 

\subsection{What $F \wedge F \neq 0$ Implies}

The six dimensional action describing the world volume dynamics of the M5-brane
contains a term which goes like $\int dB \wedge A$.  As a result, electric components of
the spacetime supergravity three-form A source a flux for the two-form B on the 
fivebrane. 

A worldvolume flux does not arise in the holomorphically wrapped 
M5-branes considered earlier \cite{Me} for the simple reason that in all these configurations 
the three-form A had purely magnetic indices. This was due to the fact that
$F \wedge F$ vanished and the equations of motion for the field strength reduced to 
$d*F = 0$. For the fivebrane system (\ref{mmaker})
under study in this note however, the situation is somewhat different. Since $F \wedge F$ is no 
longer zero, the equations of motion for F are non-trivially satisfied only when $F$ has electric 
components as well as magnetic. 

So, as a result of these Chern-Simons contributions to the equations of motion of its field strength, 
the supergravity solution for the $M5 \perp M5 (1)$ system contains contains electric 
four-form components $F_{01M{\bar N}}$ and $F_{01My}$ which, 
through the worldvolume coupling described above, source a two-form gauge field on the fivebrane. 

\subsection{A Rule And An Exception}

In our attempt to provide a self-consistent explanation of the situation, we will now 
approach the  $M5 \perp M5 (1)$  system from another point of view and present 
a different argument for the existence of a world-volume two-form. 

Consider a $p$-brane which has a $q$ dimensional intersection 
with another $p$-brane. Since the two branes have the same dimension,
this is referred to as a 'self-intersection'. From the point of view of the 
$(p+1)$ dimensional worldvolume theory, the intersection corresponds 
to a dynamical object only there is a $(q + 1)$ form to which it can couple. 

All $p$-branes contain scalar fields $\phi$ which describe their transverse
motion. The 1-form field strength of these scalars $F_1$ 
is the Hodge dual (on the worldvolume) of the $p$-form field 
strength $F_p$ of a $(p-1)$ form gauge field 
$A_{p-1}$, i.e $$ d \phi = F_1 = * F_{p} = * d A_{p-1}$$
Since the gauge field $A_{p-1}$ couples to an object with $(p-2)$ spatial directions,  
a $p$-brane can have a $(p-2)$ dimensional dynamical self intersection
\cite{p-2} .

Hence, supergravity solutions corresponding to multiple 
othogonally intersecting M5-branes can be constructed simply by 
ensuring that the branes are aligned in such a way that each 
pair of M5-branes has a 3 dimensional spatial intersection. 
So when two M5-branes intersect along a string, it appears that the 
intersection will be non-dynamical and for this reason, the intersection
$M5 \perp M5(1)$ was declared to be forbidden, or problematic, or an overlap rather 
than an intersection; at all events, the consensus seemed to be that this was a system best
 left alone. 

However, a closer inspection shows that these problems might be the result of our somewhat 
restrictive assumptions. In the derivation of the $(p-2)$ rule we have assumed implicitly 
that the only fields present in the world-volume theory are scalar fields.  If we relax this 
assumption, the one-dimensional intersection of two M5-branes can be made dynamical by 
turning on a two-form B on the worldvolume. In the presence of such a two-form, there 
seems to be no reason to rule out the possibility of the previously forbidden intersection 
$M5 \perp M5(1)$.  

An alternate 'derivation' of the self-intersection rule is based on the 
BPS $\rightarrow$  no-force argument \cite{noforce}.  Orienting branes of the 
same type so that they exert no force on each other, as must be the case for 
stable configurations, it is found that each pair of branes must share $(p-2)$
spatial directions. Applying 
this logic to the case it hand, it would appear that two M5-branes, oriented
such that M5 $\perp$ M5(1), would exert a non-zero force on each other. The
system can only be made stable if a flux is introduced to cancel this force. While an 
explicit calculation still needs to be carried out, we propose that the flux needed
to stablise the system is precisely the flux supplied by the two-form B.

\subsection{The Membrane Connection}

The only other M-brane configuration in which the constituent branes have a 
string intersection is $M2 \perp M5(1)$. Because the two M-branes have different dimensions,
this system is exempt from the  $(p-2)$ rule and there is no apparant conceptual problem 
with the existence of such an intersection. 

A link between the configurations $M2 \perp M5(1)$ and $M5 \perp M5(1)$ was 
hinted at in \cite{TheG} where it was conjectured that 
a pair of M5-branes overlapping in a string was in fact the limiting case of a 
membrane-fivebrane system, in the limit where the membrane stretched 
between two overlapping M5-branes has collapsed. This interpretation provided us with a 
way to avoid confronting the problematic $M5 \perp M5(1)$ intersection directly. 
Rather than treating 
it as an entity in itself, we dealt with it only as one particular limit of a system built 
up from a pair of the already known  and understood $M2 \perp M5(1)$ configurations.   

In light of the discussions presented above, we propose that the $M2 \perp M5(1)$ and 
$M5 \perp M5(1)$  system {\bf are}  in fact linked, and that the real link between these 
configurations lies in the  two-form which lives on the worldvolume of the fivebrane in both 
cases. Since it was already known that a worldvolume two-form must exist on the fivebrane  
in order for the intersection $M2 \perp M5(1)$ to be dynamical, we suggest
the presence of an intermediate membrane in the $M5 \perp M5(1)$ system was only 
postulated in an effort to explain the presence of three-form flux 
in a purely fivebrane configuration. 

Though this flux was not clearly visible in the harmonic function rule analysis \cite{monster} of the 
$M5 \perp M5 (1)$ system, its existence behind the scenes was manifested in certain 
strange effects -- the lack of dependence on the overall transverse direction $y$, for instance. 
By adding a membrane for which $y$ is a world-volume coordinate, we were able to explain 
away the $y$-independence of the supergravity solution.   The analysis carried out here, on the
 other hand, makes the background flux transparent and
we are able to obtain explicit expressions for electric components of the four-form field 
strength. We see that $M5 \perp M5(1)$ is a perfectly sensible system on its own, as long 
as a two-form lives on its worldvolume, and that it is not necessary to
resort to the introduction of a membrane in order to make this configuration consistent. 
Flux for a two-form gauge field is all that is needed and this is perfectly capable of existing 
even in the absence of a membrane. 

\section{The Hydra-ness of It}

In some ways, this M5-brane system is rather like Hydra, the monster in Greek mythology 
which grew two heads when one was chopped off; every question we attempt to answer 
leaves many new questions in its stead. 

The most obvious question facing us now is to find more general solutions of the supersymmetry
preservation conditions, with additional components of the four-form turned on.  In particular, by
allowing $F_{MNPy}$ to be non-zero we can turn on $F_{suvw}$ and modify the  
expression for $F_{MN\bar{P}\bar{Q}}$, leading to a far more 
complicated structure for the field strength than has been considered in this note. 

Another avenue to explore is the dimensional reduction of the M-Theory system to an 
intersecting brane configuration in Type IIA and IIB.  We would expect the issues that show up 
in the M5 $\perp$ M5(1) to have analogues in ten dimensions and 
indeed this is so. As an example, consider the following.  Beginning with (\ref{mmaker}), 
reducing along $X^1$  and relabelling the coordinates, we obtain the following IIA configuration.
\be
\begin{array}[h]{|c|c|cccccccc|c|}
  \hline
   \; & 0 & 1 & 2 & 3 & 
              4 & 5 & 6 & 7 & 8 & 9\\
  \hline
  {\bf D4} & \times & \times & \times & \times & \times
               &  &  &  &  &  \\
  {{\bf D4}} & \times &  & 
               &  &  & \times & \times & \times & 
\times & \\
  \hline
\end{array}
\ee
T-Dualities along $X^1 X^2 X^3$ and $X^4$ take us to 
\be
\begin{array}[h]{|c|c|cccccccc|c|}
  \hline
   \; & 0 & 1 & 2 & 3 & 
              4 & 5 & 6 & 7 & 8 & 9\\
  \hline
  {\bf D0} & \times &  &  &  & 
               &  &  &  &  &  \\
  {{\bf D8}} & \times & \times & \times
               & \times & \times & \times & \times & \times & 
\times & \\
  \hline
\end{array}
\ee
It was shown in \cite{IIA} that this brane system is stable and supersymmetric 
only when a background flux is turned on.  Recall that all the arguments presented 
in this note have converged around a central issue: {\sf  \bf non-zero electric components 
for the four-form field strength are needed in order for M5 $\perp$ M5(1) to be an allowed 
intersection}.  We suggest that the flux needed to stabilise the D0 $\perp$ D8 brane configuration in 
Type IIA is simply the lower dimensional (compactified) manifestation of these four-form 
components. 

Many intersecting brane systems can be obtained in string theory by reducing 
the M-Theory system (\ref{mmaker}) to ten dimensions and performing a series of 
dualities. We list some examples below.
\be
\begin{array}[h]{|cc|lllll|}
\hline
& &  D4\perp D4(0) & \; & D6\perp D2(0) & \; &  D8 \perp D0 (0)\\
{\bf \sf Type \; IIA} & & & & & & \\
& & NS5 \perp D4 (1) &  \; &  NS5 \perp D6 (3) & \; &  NS5 \perp D8 (3) \\
\hline
 & & 
 D5 \perp D3 (0) & \; &  D7 \perp D1 (0) &  &\\
{\bf \sf Type \; IIB} & & & & & & \\
& & NS5 \perp D3 (0) & \; &   NS5 \perp D6 (2) & \; & NS5 \perp D7 (4) \\
\hline
\end{array}
\ee
As they stand, we expect these intersecting brane systems to be unstable and/or 
non-supersymmetric. It should, however, be possible to make
these systems stable by turning on a suitable background flux and a purely string 
theoretic calculation should enable us to work the flux needed.  On the other hand, if we have traced 
the origins of the stability criterion correctly, it should also be possible
to arrive at the same result by starting out from M-Theory.  If we follow the 
four-form field strength for the M-brane system (\ref{mmaker}), tracing it through 
compactification and the relevant dualities needed to arrive at a particular 
string theory system, we should reproduce the background flux which string theory 
says is needed to stablise the ten dimensional brane configuration arising at the end of 
the duality chain. Performing these calculations and verifying this conjecture is another 
task we have ahead of us.  

Yet another question which naturally springs to mind is the connection of the M-brane 
system (\ref{mmaker}) to calibrations. An extension of the concept of calibrations \cite{GCal} has 
been proposed for branes with non-trivial world-volume fields \cite{calibration}. It would be 
interesting to explore this further\footnote {Just as this paper was being submitted, reference 
\cite{justnow} appeared which also discusses M5-branes with a worldvolume two-form, in the
 context of generalised calibrations.}  and attempt to find calibrated forms in the background 
generated by holomorphic M5-brane wrapped on a four-cycle in $\C^4$.

So, one head of this monster has been cut down only to be replaced by many others.  
To avoid feeling overwhelmed, remember that we had to begin the battle somewhere! The little 
skirmish outlined above has helped us gain a feel of the way things work in this M-brane system.  
Although we are not claiming to have won a decisive victory yet, we are confident that we have 
learnt a lot about the enemy and that our knowledge will prove useful in the future.  
For now, we will continue our quest, hoping that a day will dawn when this Hydra finally 
runs out of heads. 

\acknowledgments{
I am indebted to Jerome Gauntlett, George Papadopoulos, Arkady 
Tseytlin and Paul Townsend for very interesting and enlightening 
discussions. I would also like to express my gratitude to Ansar 
Fayyazuddin, who has been there from the beginning and suffered 
patiently through the sometimes painful development of this 
project from a single intractable equation to the full-blown 
problem it is now! I am also grateful to Dileep Jatkar, for a critical reading
of the draft and conversations which were as much fun as they were useful. Special
thanks to Midhat Kazim, for his encouragement, interest and comments on the manuscript.}


\newpage

\begin{thebibliography}{99}
\bibitem{Me}
T.Z.Husain, ``{\it M2-branes wrapping Holomorphic Curves}'', 
\hepth{0211030}\\
T.Z.Husain, ``{\it That's a Wrap!}''
\hepth{0302071}
%
\bibitem{Julie}
J. Blum, ``'{\it Triple Intersections and Geometric Transitions}'
\hepth{0206248}
%
\bibitem{p-2}
A. Strominger ``{\it Open P-Branes}, \hepth{9512059}\\
P.K. Townsend ``{\it Brane Surgery}'' \hepth{9609217}
%
\bibitem{noforce}
A.A. Tseytlin,       
``{\it No-force condition and BPS combinations of p-branes in 11 and 
10 dimensions}'' \hepth{9609212}.
%
\bibitem{Tseytlin}
 A.A. Tseytlin, ``{\it Harmonic superpositions of M-branes}''
\hepth{9604035}
%
\bibitem{LecNotes}
G.Papadopoulos and P.K.Townsend,``{\it Intersecting M-branes}''
\hepth{9603087}\\
K.S.Stelle, ``{\it BPS Branes in Supergravity}''
\hepth{9803116}.\\
J.M.Figueroa-O'Farril, ``{\it Intersecting Brane Geometries}''
\hepth{9806040}.\\
P.K.Townsend, ``{\it Brane Theory Solitions}'', 
\hepth{0004039}\\
D. J. Smith, ``{\it Intersecting Brane Solutions in String and 
M-Theory}'' \hepth{0210157}.\\
T.Z.Husain, ``{\it If I Only Had A Brane!}''
\hepth{0304143}\\
%
\bibitem{monster}
J. P. Gauntlett, D. A. Kastor and J. Traschen,
``{\it Overlapping Branes in M-Theory}'' \hepth{9604179}\\
A.A.Tseytlin, 
``{\it Composite BPS configurations of p-branes in 10 and 11 dimensions}'' 
\hepth{9702163}.

\bibitem{FS}
A.Fayyazuddin and D.J.Smith, ``{\it Localised Intersections of 
M5-branes and Four-dimensional Superconformal Field Theories}.''
\hepth{9902210}\\
B.Brinne, A.Fayyazuddin, D.J.Smith and T.Z.Husain, 
``{\it N=1 M5-brane Geometries}''
\hepth{0012194}
%
\bibitem{BBS}
K.Becker, M.Becker and A.Strominger, ``{\it Fivebranes, Membranes and 
Non-perturbative String Theory}''
\hepth{9507158}.
%
\bibitem{wrapM5}
A. Fayyazuddin and M. Spalinski
``{\it Extended Objects in MQCD}''
\hepth{9711083}\\
%
\bibitem{TheG}
J. P. Gauntlett, ``{\it Intersecting Branes}'' \hepth{9705011}
%
\bibitem{IIA}
O. Bergman, M.R. Gaberdiel, G. Lifschytz
``{\it Branes, Orientifolds and the Creation of Elementary Strings}'', 
\hepth{9705130}\\
M. Mihailescu, I.Y. Park and T.A. Tran
``{\it D-branes as Solitons of an N=1, D=10 Non-commutative Gauge Theory}''
\hepth{0011079}\\
E.Witten, ``{\it BPS Bound States Of D0-D6 And D0-D8 Systems In A B-Field}''
\hepth{0012054}
%
\bibitem{GCal}
J.P.Gauntlett, N.D.Lambert and P.C.West,
``{\it Branes and Calibrated Geometries}''
\hepth{9803216}.\\
G.W.Gibbons and G.Papadopoulos, ``{\it Calibrations and Intersecting
 Branes}'',\hepth{9803163}.\\
J.Gutowski and G.Papadopoulos ``{\it AdS Calibrations}'', \hepth{9902034}\\
J.Gutowski, G.Papadopoulos and P.K.Townsend, ``{\it Supersymmety and Generalised 
Calibrations}'', \hepth{9905156}\\
J.P.Gauntlett, ``{\it Branes, Calibrations and Supergravity}'', \hepth{0305074}
%
\bibitem{calibration}
O. Baerwald, N. D. Lambert, P. C. West, 
``{\it A Calibration Bound for the M-Theory Fivebrane}''
\hepth{9907170}
%
\bibitem{justnow}
D.Martelli and J.Sparks, ``{\it G-Structures, Fluxes and Calibrations in 
M-Theory}'' \hepth{0306225} 
\end{thebibliography}
\end{document}